\documentclass{aa}

\usepackage[dvips]{graphicx}
\usepackage{longtable}
\usepackage{amssymb}

\begin{document}

\title{Automated Spectroscopic abundances of A and F-type stars\\ using echelle
  spectrographs }

\subtitle{I. Reduction of ELODIE spectra and method of abundance
  determination \thanks{Based on
    observations collected at the 1.93\,m
    telescope at the Observatoire de Haute-Provence (St-Michel
    l'Observatoire, France)}}

\author{D. Erspamer \and P. North}

\offprints{D. Erspamer}

\institute{Institut d'astronomie de l'Universit\'e de Lausanne, 
              CH-1290 Chavannes-des-Bois, Switzerland}

\date{Received 11 October 2001 / Accepted 12 November 2001}

\abstract{This paper presents an automated method to determine
  detailed abundances  
  for A and F-type stars. This method is applied on spectra taken with 
  the ELODIE spectrograph. Since the standard reduction procedure of
  ELODIE is optimized to obtain accurate radial velocities
  but not abundances, we present a more appropriate reduction
  procedure based on IRAF. We describe an improvement of the method of 
  Hill \& Landstreet (\cite{hill}) for obtaining $V\sin{i}$,
  microturbulence and 
  abundances by fitting a synthetic spectrum to the observed one. In
  particular, the method of minimization is presented and 
  tested with \object{Vega} and the \object{Sun}. We show that it is
  possible, in the case of the \object{Sun}, to recover the abundances of
  27 elements well within 0.1 dex 
  of the commonly accepted values.
}

\titlerunning{The method with ELODIE spectrograph}

\maketitle

\section{Introduction}

The determination of detailed abundances requires a high resolving
power ($> 30\,000$) 
and a wide spectral range. In order to satisfy both
requirements simultaneously, echelle spectrographs must be used.
ELODIE (Baranne et al.
\cite{baranne}, hereafter BQ96) is a fiber-fed
echelle spectrograph with a resolution of R=42\,000 attached to the 1.93\,m
telescope of the Observatoire de Haute-Provence (OHP), France. This
spectrograph and its reduction software were optimized to measure
accurate radial velocities.

In this paper we first show what precautions have to be taken to use ELODIE
for other spectroscopic analyses, in our case detailed abundance
determinations. To achieve our goal we had to make another reduction,
starting from the raw image and taking special care in the removal
of scattered
light. Another important point in the reduction is to paste together the
different orders of the spectrum and 
normalize them. Secondly, we present a method to estimate 
abundances with synthetic spectra adjustments. This method is an
improvement of 
that of Hill \& Landstreet (\cite{hill}, HL93 hereafter). 
It is automated as much as possible and is able to analyse stars with
various rotational velocities (up to 
150 $\mathrm{km~s^{-1}}$), for which the 
equivalent width method is not applicable.
Finally, to assess the validity of this 
method, we compare the abundances derived for 
\object{Vega} (\object{$\alpha$ Lyr} = \object{HR 7001} = \object{HD 172167})
and the \object{Sun}
with those in the literature. These two reference stars are used to
check the code's validity for stars having effective temperatures
between those of the \object{Sun} and of \object{Vega}. 

Analysis tools with related goals but different approaches have been
developed by Valenti and Piskunov (\cite{valenti}), Cowley
(\cite{cowley}) and Takeda
(\cite{takeda}). Takeda's method has been used by Varenne and Monier
(\cite{varenne}) to
derive abundances of A and F-type stars in the Hyades open cluster.

\section{Observations}\label{obs}

The spectra used in this work were obtained with the ELODIE echelle
spectrograph (see BQ96 for technical details) attached to the 1.93\,m
telescope of the Observatoire de 
Haute-Provence (France) during August 1999. A high S/N ($ > 300$)
spectrum of \object{Vega} was obtained covering the range of
3900-6820\AA. 

The best way to
check both reduction and analysis is to obtain a good quality
solar spectrum with ELODIE, in the same
conditions as in the case of stellar observations. The target can be 
either an asteroid or one of the Jovian moons in order to have a
point-like, but bright enough source. Our choice was
\object{Callisto}, and it was observed on 13 August 1999 when 
it was almost at a maximum angular distance from Jupiter, 
thus avoiding light pollution by the 
planet. The resulting spectrum has a S/N of about 220 at 5550\AA. 

\section{Data reduction}

The primary goal of the ELODIE spectrograph was to measure high-accuracy
radial velocities, and 
the data reduction pipeline was optimized for that purpose. The
on-line reduction
is achieved by the software INTER-TACOS (INTERpreter for the Treatment,
the Analysis and the COrrelation of Spectra) developed by D. Queloz and 
L. Weber at Geneva Observatory (BQ96). During
this reduction, the background is removed using a two dimensional
polynomial fit that has a typical error of about 5\%, with a peak in
the middle of the orders (cf. Fig.~11 of BQ96). We tried
to improve this fit by increasing the 
polynomial order. However, we encountered an internal dimensional
limitation that prevented us from using a high enough order to correct the
middle peak. Therefore, we decided to use IRAF (Image Reduction and
Analysis Facility, Tody, \cite{tody} ). Another point motivating  our
choice was the wide availability of IRAF.

\subsection{Overview of the reduction procedure}

The reduction
itself was done with IRAF, and more precisely with the \texttt{imred.ccdred}
and \texttt{imred.echelle} package. Its main stages are the following :

\begin{itemize}
\item Combination and averages of the calibration exposures (offsets, darks,
  "flat-fields") with the IRAF functions \texttt{zerocombine, darkcombine} 
  and \texttt{flatcombine}.
\item Correction of the object exposure from the offset, dark and bad pixels,
  taking into account the overscan level with
  \texttt{ccdproc}.
\item Finding, centering, resizing (i.e. function that estimates the
  width of the orders), and tracing the orders using the
  flat image with \texttt{apfind, apcenter, apresize, aptrace}.
\item Removal of the diffuse light of all raw images except the
  wavelength calibration one with \texttt{apscatter}.
\item Recentering, resizing, retracing and extracting the
  \textit{flat-field spectrum} with \texttt{apsum}.
\item Extraction and calibration of the thorium image with
  \texttt{apsum, ecidentify, ecreidentify}.
\item Extraction of the object spectra using the variance weighting
  method (Horne \cite{horne}) with \texttt{apsum}.
\item Division by the extracted \textit{flat-field spectrum} with \texttt{sarith}.
\item Wavelength calibration of the object spectra with
  \texttt{dispcor}.
\item Merging the 67 orders with a home-made fortran program to obtain one
  spectrum for the whole range of ELODIE.
\item Normalization to the continuum using the IRAF \texttt{continuum}
  function. 
\end{itemize}

\subsection{Removal of scattered light}

\begin{figure}[tbh]
\resizebox{\hsize}{!}{\includegraphics{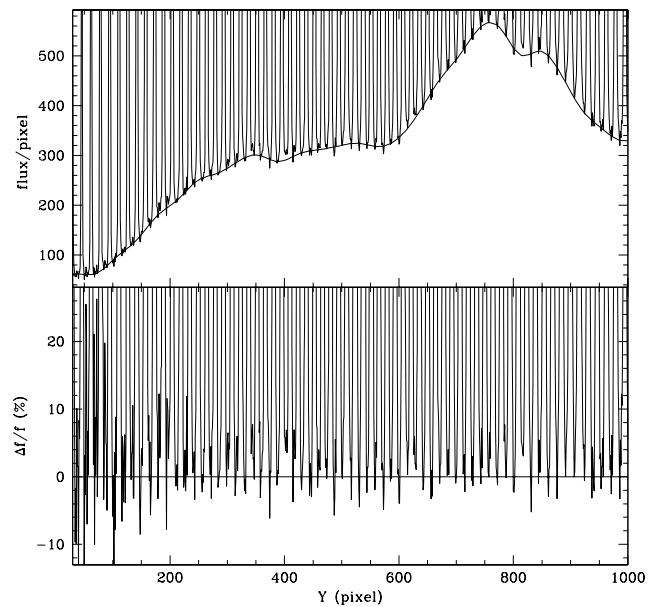}}
\caption{\textbf{Top :} cross order tracing at X=512 of a localization 
  exposure superimposed with the fit of the
  background. \textbf{Bottom :} difference between the fit and the
  background level $\Delta f/f = (f_{image} - f_{fit})/f_{fit}$. The
  typical error in the background measurement is below 2\%.}
\label{lumdiff}
\end{figure}

The main weakness of the online procedure resides in the background
subtraction. Although a typical error in the background measurement of 5\%
can be tolerated for accurate radial velocity measurements (BQ96), it is
important to achieve a better adjustment in 
order to use the spectra for abundance measurements.
The scattered light is estimated, smoothed and subtracted in
  the following way. Interorder pixels are fit with a one dimensional
  function in the direction perpendicular to the dispersion. This
  fitting uses an iterative algorithm to further reject   
  high values and thus fits the minima between the spectra. The fitted
  function is a combination of 30 spline functions (see
  Fig.~\ref{lumdiff} Top). Because each fit (each column) is done
  independently, the scattered light thus determined will then be
  smoothed by again fitting a one dimensional function (30
  spline functions in the dispersion direction). The final scattered
  light surface is then subtracted from the input image. The reason
  for using two one-dimensional fits as opposed to a surface fit
  is that the shape of the scattered light is generally not easily modeled
  by a simple two dimensional function.
 The typical error in
the background measurement is about 2\%. This is shown in
Fig.~\ref{lumdiff}, which should be compared with Fig.~11 of BQ96. 

It is to be noted that the blue orders are not very well
corrected. However, this is a deliberate choice. We cannot
simultaneously adjust the first orders without using more than 35 cubic 
spline functions. But with that number, the fitting function is too
sensitive to the order. Moreove, since the signal to noise ratio (S/N)
is lower in the bluest orders (see Fig.~10 of BQ96), these are not optimal
for abundance determination. In these orders, it is very difficult
to adjust 
the continuum because of the calcium and Balmer lines. Therefore, we
decided not to use the first orders and the problem of the background
subtraction in them was left unresolved.

\subsection{Minor changes}
During the observing run every night, we did many offsets,
darks and flat-fields. Instead of using only the last exposure for the
offset, dark and 
flat-field correction, as is the case in the online reduction, 
we chose to combine the exposures in order to
remove pixels hit by cosmic rays, and obtain a mean offset, dark and
flat-field. Then we corrected each pixel 
of the object image with the corresponding one of the offset and
dark. 

Then we used the average flat-field (while the standard TACOS reduction 
only uses the last one) to determine the shape of the
orders and this shape was used as reference for the extraction of
the object image. We took care to adjust the resizing parameter with the lowest
possible parameters in order to get almost all the flux. Finally, we
set the aperture limit at 0.005 time the level of the peak. This lead 
to the extraction of 99.9\% of the flux spread over the order.

As explained in the paper BQ96, the \textit{flat-field spectrum}
correction method (i.e. flat-field correction after extraction of the
spectrum) is satisfactory with such a stable instrument as ELODIE. This
method is also applied in our reduction (in any case, it is not really 
possible to get a true \textit{flat-field image} with ELODIE).

The wavelength calibration is carried out using the thorium spectrum. The
spectra are extracted without correction of the scattered light and without
the flat-field division. A two dimensional Chebyshev polynomial is used to
constrain the global wavelength position with the degree 7 for both
directions. The typical rms between the fit and the location of the
lines is always below 0.001 \AA for the wavelength
  calibration of the whole spectrum. The fit is just a formal one. 
  We did not attempt to model the optical behaviour of the
  spectrograph. We used the thar.dat file from IRAF to identify the 
  lines. This file contain the line list of a Thorium-Argon spectral 
  Atlas done by Willmarth and collaborators available at 
  http://www.noao.edu/kpno/specatlas/thar/ thar.html which used 
  identification from Palmer and Engleman (\cite{palmer}, the same as
  BQ96) for Th and
  from  Norl\'en (\cite{norlen}) for Ar. Looking carefully at the flux
  ratio in Fig.~\ref{compare}, bottom, a number of the
  larger discrepancies appear to be due to minute wavelength
  differences between both spectra. A difference of 
  $50\,\mathrm{m~s^{-1}}$ might already
  explain such a signature in the ratio panel.
  Fig.~14 of BQ96 shows that the accuracy differs from one order to
  the other.
  Fig.~\ref{compare} displays more than two orders, and differences appear
  only in the left and right parts, which correspond to 
  different orders than 
  the central part. As accuracies are different, it is
  possible that small shifts exist between orders.

\subsection{Merging the orders}
\begin{figure}[tbh]
\resizebox{\hsize}{!}{\includegraphics{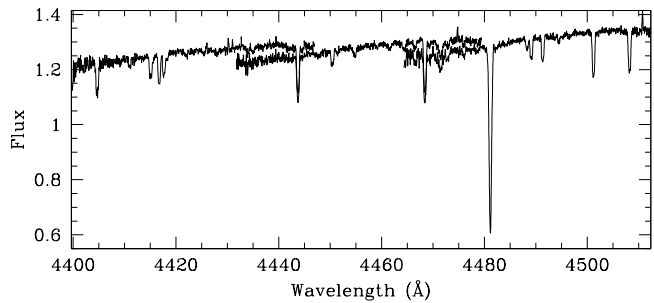}}
\caption{19th, 20th and 21th orders of \object{Vega} before merging}
\label{recol}
\end{figure}

The next important task is to merge the orders to obtain a one
dimensional spectrum covering the whole wavelength domain. At that
point, we encountered a problem with the data. The extracted orders are
not flat enough to be merged using the average or median value of the
order as coefficient (see Fig.~\ref{recol}).

\begin{figure*}[tbh]
\centering
 \includegraphics[width=17cm]{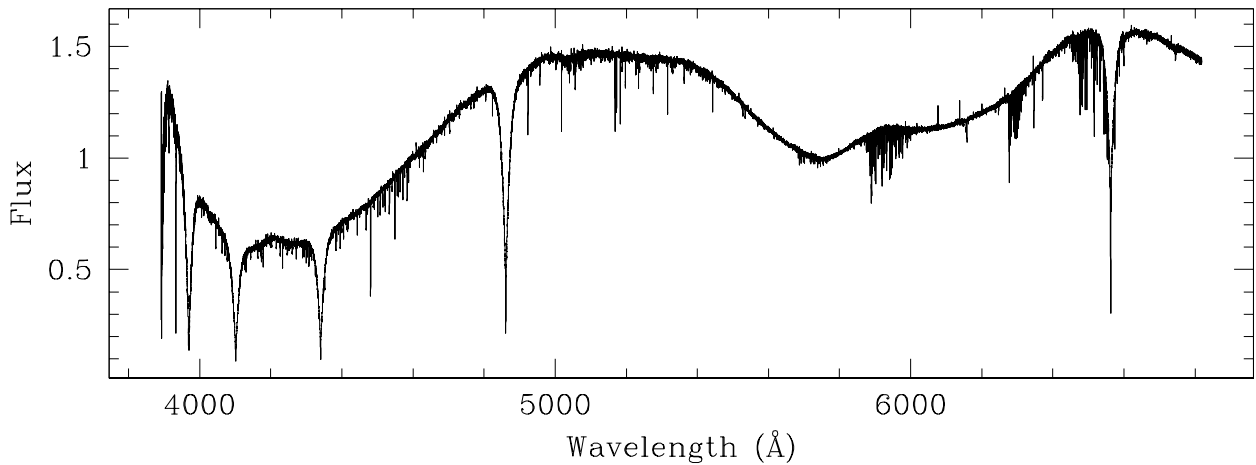}
 \caption{The whole spectrum of \object{Vega} }
 \label{vega}
\end{figure*}

Merging by considering only one average value per order
results in a spectrum with 
steps (imagine Fig.~\ref{recol} with a vertical line connecting the
middle of the overlapping region, and smooth that transition region
over 10 pixels).

We decided to compute our own
program to paste the orders. There is an overlapping
  region until the 64th order. However the overlapping
region is large enough to estimate the ratio only until the 50th order. 
(Note that the orders are numbered from 1 to 67 and the
``true'' number in not used as in BQ96). Therefore, we used two
different merging methods, one using 
the overlapping region for the orders 1 to 50, and another using the
first and last 200 points of the order (each order is rebinned with a
step of 0.03\AA~before the merging). With both methods, we computed a
ratio allowing to scale the orders, starting from the middle order
which is used as reference.

In the first method, we computed the average of the ratios of the
overlapping points 
and the rms scatter. Then, we did a loop taking into account only the
ratios between the average $\pm 2 \sigma_{rms}$ until no
points were deleted or the number of points become $\leq 50$. This
method was very efficient, and worked in almost every case. 

The second method was not quite as efficient but we rarely had to
correct its results manually. We decided to use the
first and last 200 points of the orders, compute the average value of
these points and the rms scatter, then recompute the average but
deleting the points that were not between the average $\pm 2
\sigma_{rms}$ until no 
points were deleted or the number of points became $\leq 50$ and
finally compute the ratio of the averages of the end of an order and
the beggining of the following order.

Finally, starting from the middle order, the orders are scaled
  by multiplicative adjustments. In the overlapping regions, no
  attempt was made to make a weighted average : in view of the blaze
  function, it was decided to retain the flux of the first order for
  3/4 of the overlapping region and the flux of the following order
  for the remaining 1/4.
Both methods are compatible and it is possible to merge all orders
in a single pass; Fig.~\ref{vega} shows the results for \object{Vega}.

\subsection{Normalization}

\begin{figure}[tbh]
\resizebox{\hsize}{!}{\includegraphics{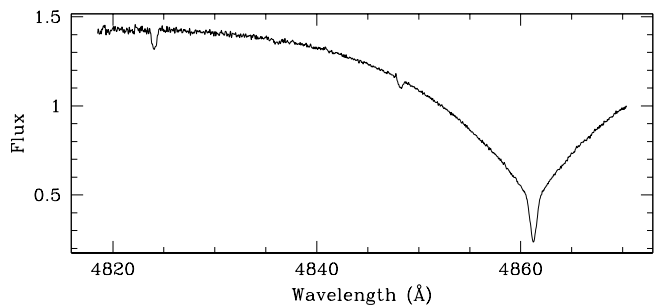}}
\caption{Plot of the 31th order of \object{Vega} around H$\beta$}
\label{hbeta}
\end{figure}

The final step is normalization to the continuum level. A simple
look at Fig.~\ref{vega} shows that it is no easy task, especially
around the Balmer lines H$\alpha$ and H$\delta$ and the \ion{Ca}{ii}~K 
line. We decided to use the function \texttt{continuum} of IRAF. However,
it is very hard to normalize the whole spectrum in a row. One could argue 
that, if the normalization was done before merging, that operation would
become much easier. However, some orders are not normalizable,
especially those  
containing the Balmer lines (see Fig.~\ref{hbeta}).

We chose to split the whole spectrum into 6 parts, and normalize each
part separately (besides analyzing the whole spectrum at once would require
too much data processing). The task \texttt{continuum} has many
parameters and the result 
is very dependent on them. However, once a good set of parameter is
defined, it can be used for a lot of different spectra. Moreover,
IRAF allows to modify the parameters interactively in case of
unexpected behaviour. 

Although IRAF works well automatically, it is important to check 
all the spectra visually. Unfortunately, despite all different
numerical tests, the eyes 
appear to be still the best way to decide which set of parameters
to use.

\subsection{Check with a reference spectrum}

\begin{figure*}[tbh]
\centering
 \includegraphics[width=17cm]{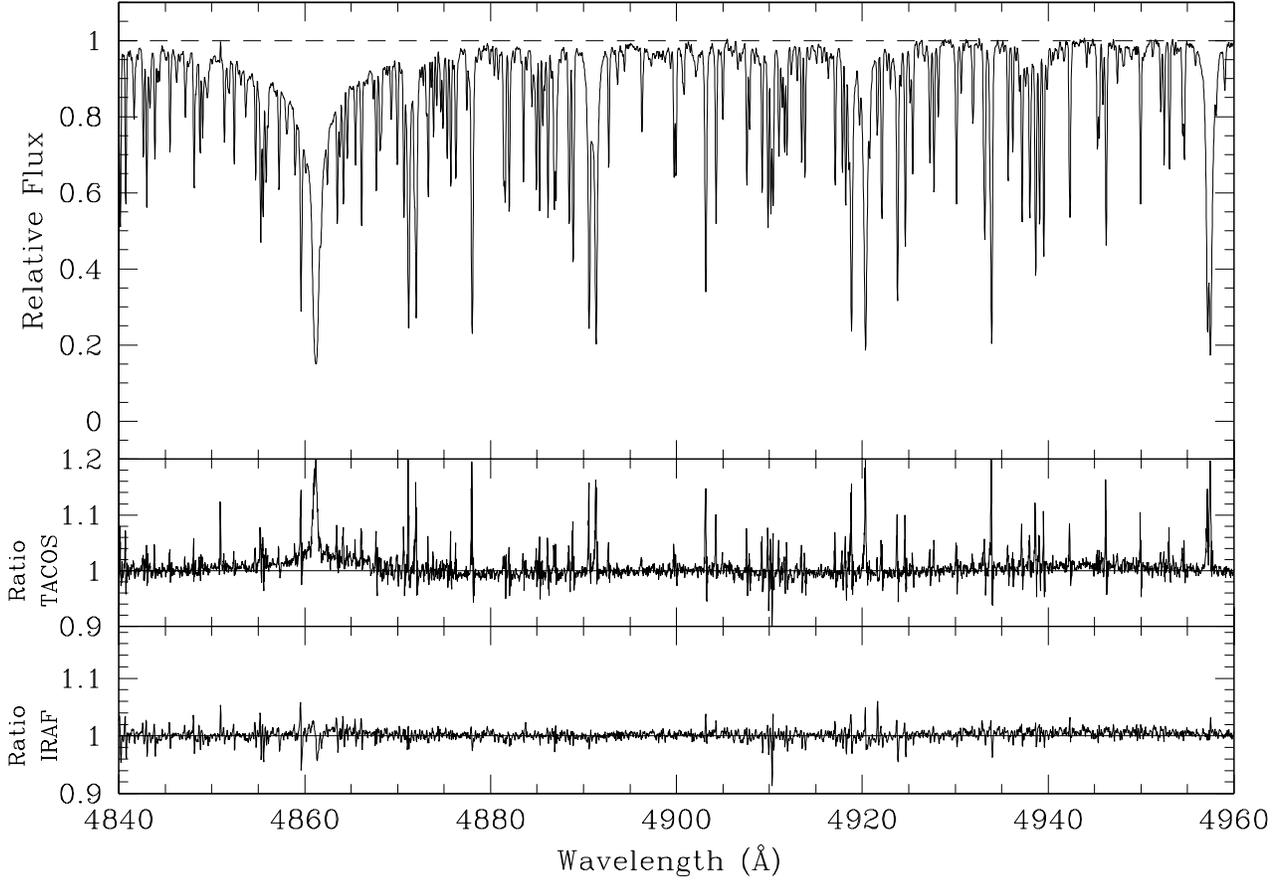}
 \caption{\textbf{Top :} Solar spectrum extracted
   with the optimized IRAF reduction. \textbf{Middle :} Ratio between the
   spectrum resulting from the standard TACOS procedure and the solar
   spectrum from Kurucz 
   \textbf{Bottom :} Ratio between the spectrum resulting from the
   optimized IRAF reduction and the solar spectrum from Kurucz }
 \label{compare}
\end{figure*}

Our reduction was checked using the Solar Atlas from
Kurucz et al. (\cite{kurucz}). This spectrum was acquired with a very high
resolving power ($ 300\,000$) and a very high signal to noise ratio
($ 3\,000$). The resolving power was adjusted to
that of ELODIE by convolving the
spectrum with an instrumental profile; a simple gaussian with an FWHM
corresponding to the normal resolution R=42000 was considered sufficient.
Our comparison spectrum was acquired using \object{Callisto} so that
we were in a stellar-like configuration. This precaution is not very
important as ELODIE is a fiber-fed spectrograph, but one of the advantages
was that it required a rather long exposure and therefore the reduction was
sensitive to the dark correction. Finally, we adjusted the radial
velocities. Notice that two versions of the spectrum, one resulting from
the TACOS reduction procedure and the other from the IRAF procedure,
were merged and normalized using our method. The
comparison is illustrated in Fig.~\ref{compare}. 

It is clear, looking at the ratio for the strong lines, that  scattered
light is not well subtracted with the standard TACOS procedure. The
difference increases as lines strengthen, reaching a maximum at the core 
of H$\beta$ in our example. Even if the difference for
H$\beta$ can partly come from the normalization as can be seen looking
at  the ratio in the wings, which differ slightly from 1, the big
difference in the core cannot be assigned to different continuum
adjustment. On the contrary, our optimized reduction leads to
differences which remain within, or only slightly larger than the noise.

\section{Abundance analysis}

This section presents the method for abundance analysis. In
the first part (Sect.~\ref{synspec}), the spectrum synthesis
program is described. In the second (Sect.~\ref{abund}), the
minimization method is explained. 

This method adjusts the
abundances using synthetic spectra. The starting point was the program 
described in Hill \& Landstreet (\cite{hill}), which was used to determine
detailed abundances in A-type stars and has been kindly provided by
Dr. G.M. Hill. The modifications made to this
program will be presented in this section.

\subsection{spectrum synthesis}\label{synspec}

The spectral synthesis code used here is similar to the one described in
HL93. It is an LTE synthesis code (see HL93 for details). Various
modifications were done :

\begin{itemize}

\item
The code from HL93 used the model atmospheres ATLAS5 of Kurucz
(\cite{kurucz2}). It was modified to allow the use of ATLAS9
models. However, the format of the models has changed. In ATLAS5,
there was a column containing the geometrical depth X and another
containing the optical depth at 5000\AA, $\tau_{5000}$, and the
code of HL93 uses it to compute the optical depth at each wavelength of
the spectrum ($\tau_{\lambda}$). However, in ATLAS9 the depth is only
given in the 
``rhox'' parameter which is the
density integrated as far as the geometrical depth X. In order to modify
the code as little as possible, we only changed the way
$\tau_{\lambda}$ is computed. A subroutine of the ADRS code
(Y. Chmielewski, \cite{chmiel}) was used to reformat the model as a
function of $\tau_{5000}$. 
Then it was easy to use $\tau_{5000}$ instead of X and the following
equation was used for the calculation of the optical depth at each
wavelength step.  
\[  \tau_{\lambda} = \int{\frac{\kappa_{\lambda}}{\kappa_{5000}}
  d\tau_{5000}} \]
ATLAS9 was installed on a SunBlade100 computer (Sun UltraSparcII
  machine). The Unix version adapted by M. Lemke
\footnote{http://www.sternwarte.uni-erlangen.de/ftp/michael/
atlas-lemke.tgz}  was used
because the version of Kurucz (\cite{kurucz3}) is adapted to 
VAX-VMS. It will allow to compute models with the right effective
temperature, gravity, metallicity and microturbulent
velocity. Models will be computed without overshooting as it appears
that they are best adapted to abundances
analysis (see for example Castelli et al., \cite{castelli})

\item
The second modification allows the use of the VALD database (Piskunov
et al. \cite{vald1}, Kupka et 
al. \cite{vald2} and Ryabchikova et al. \cite{vald3} )  as the source
of line list parameters. This 
database provides $\gamma_4 / 
N_e$ and $\gamma_6 / N_H$ at 10\,000K for the Stark and van der Waals
damping parameters respectively instead of C$_4$ and C$_6$ used by HL93.
In case these values are not available, the damping parameters are
calculated using the formulae described in HL93.

\item
The polynomial partition functions were completed for
as many elements as possible, using data from Irwin
(\cite{irwin}) and Sauval \& Tatum (\cite{sauval}). In case that the 
function is unavailable, the program uses the statistical weight of the 
lowest energy level. However, for some elements like Co and Ba, using
a partition function or the approximate value may result in abundance
changes as high as 0.2 dex.

\item 
Finally, there were several other changes, including the introduction
of dynamical memory allocation. However, the theoretical 
assumptions used in the initial code do not change. 
\end{itemize}

\subsection{Abundances determination}\label{abund}

\subsubsection{Line list}\label{list}

Instead of using a set of meticulously selected lines, the first
hope was to use the line list as it comes from the VALD database, using
the ``\textit{extract stellar}'' option. This choice was motivated by the large
spectral range of ELODIE. The idea was to use a large number of lines 
with parameters not necessarily well known, but considering a
large number, the effect of poor $\log{gf}$ should
disappear and the mean value for an element should be
correct. This idea is justified for elements of the iron peak, but not 
for elements as Si, Sr, Ba, and heavier
elements. For these elements (except
for Si), only one or a few lines are present, and it is easy to understand
 that if there are
only a few lines, the abundance is very sensitive to the line
parameters.

Although VALD-2 provides the most recent collection of oscillator
strength data, it appears that for some elements with few lines (and Si),
the $\log{gf}$ values had to be examined 
individually and adjusted whenever possible (i.e. when the line was 
not blended). To achieve good adjustment, lines of problematic elements
were checked individually and 2 methods were used to adjust the
oscillator strength using the \object{Sun} spectrum:

\begin{itemize}
\item
The line was checked for presence in the line list coming from
the SPECTRUM package of 
Gray (\cite{spectrum}, and available at
http://www.phys.appstate.edu/spec-trum/spectrum.html) that contain
adjusted $\log{gf}$ for some lines. If such was the case, and if the
value was different, this value was tried. Then if it gave good results 
it was retained. 

\item
In case of failure of the first method, the value was adjusted on the
Solar spectrum with an 
iterative procedure until the abundance was correct within 0.05 dex. 

\end{itemize}

Finally, instead of using directly the VALD result, a
reference line list covering our spectral domain was compiled, and 
used for all the stars to be analyzed. It is
to be noticed that this line list may still contain erroneous $\log{gf}$ values
because every line is not necessarily visible in the solar spectrum, and also 
because this parameter was only adjusted for problematic elements when
the line was clearly 
identifiable. Moreover, lines can be individually erroneous, but when
all lines of an element are used together, the individual errors
seems to cancel out. This is the case in particular for iron where
statistically, the abundance is correct while a lot of lines have
inaccurate $\log{gf}$ values. For other elements, it is not
  obvious that these errors should average, but the application of the
  method for the \object{Sun} seems to indicate that it is the case
  (see Sect.~\ref{sect_sun}). 

\subsubsection{Minimization method}
The program of HL93 uses the technique known as ``downhill simplex
method''. This method is certainly the best if the goal is 
``to get something working quickly'', in the sense that it does not have
strong initial constraints. Another advantage is that it does
not require derivatives. However, it is not the fastest
one. Moreover, in our case, the function to be evaluated requires
computation of a synthetic spectrum. The major improvement in time will come 
from lowering the number of spectra to be computed. Looking for
minimization methods of multidimensional functions in
\textit{``Numerical Recipes''} (Press et al \cite{numrec}), we
decided to implement a \textbf{direction set method in multidimensions} known
as \textbf{Powell's method}. This method can be summarized as
follows. Given a set of N 
directions (where N is the number of free variables), the method
performs a quadratic minimization along each direction and tries to
define a new one as the direction of the largest
decrease. We simply adapt the Numerical Recipies code by
  adding some tests to avoid looping that may
occur if the program tries to adjust lines of an element that are too
faint to be of any significance.

\textbf{Initial conditions :} 
In order to use this method an initial set of
directions has to be defined. The goal is to
adjust radial, rotational and microturbulent velocities as well as
abundances. In order to be efficient even in the first iteration,
it appears that the best choice of directions is to adjust
successively radial velocity (which is fixed when it is available from
ELODIE online 
reduction), rotational velocity,
the abundances starting from the element with the maximum of
significant lines, and finally microturbulent velocity. 
As we do not know a priori the abundance pattern, the starting
point is the solar one. Note that the solar abundances will always
refer to Grevesse and Sauval (\cite{grevesse}). The order proposed is
justified by the following example.
Let us consider a blend of two lines of different elements;
the element with the largest number of lines will be adjusted
first. As it has other lines, it is less sensitive to the blend and 
the program does not try to fill the blend by increasing the abundance 
of this element only, as it would happen if the element with only one
or two lines was adjusted first. Then
the second element is adjusted in order to fill up the blend.

\textbf{Procedure of analysis :}
It is important to retain only the lines from the reference list
that contribute to the spectrum.
Therefore lines were sorted using the 
equivalent width computed with a reduced version of the program. Only
lines with an equivalent width larger than 10 m\AA~(when computed with 
an atmosphere model corresponding to the stellar parameters and solar
abundances) are used for the first abundance determination. The results
of this first minimization are used to re-sort out 
the lines with the same equivalent width criterion. Then a second
computation is done with the new line list, using the result of the
first computation as a starting point for the velocities and
abundances. That speeds up the second adjustment

This analysing procedure allows to 
eliminate a lot of lines that are significant for 
solar abundances but are no longer visible when it comes to abundances
of the star. Conversely, it may also allow to gain lines that were 
too weak for solar abundances but are strong enough with the stellar
value.

\textbf{Speed optimization :}
During abundance analysis, a synthetic spectrum is computed at each
step of the minimization procedure. It is important to find a way
to reduce the time of analysis as much as possible.

The spectral range of ELODIE is wide (3900-6820\AA). In order to have 
the best possible abundance estimates, it is important to use the
largest number of lines i.e. the widest possible spectral
range. However it is not possible, with this 
method, to use the whole spectral range at once for various reasons :

\begin{itemize}

\item
The program does not implement the computation of Balmer lines because 
HL93 only used small spectral windows. It
only implements H$\beta$ but this 
line is just treated as a correction to the continuum. The other Balmer
lines are not computed. Only H$\beta$ was included by HL93 because the HeI
$\lambda4922$ line lies in its red wing, and it was an important line
for the stars they studied.

\item
Running the program on a spectrum from 5000 to 6500 is much slower
than adjusting 4 parts of 400\AA~successively on the same
computer. This is surprising, but may come from inadequate
programming as would be the case if looping in an array using the bad
order of indices (one index is going much faster than another).

\item
The memory used by the program has to be kept within reasonable values
because it is important to be able to dispatch the task on a large number of
computers as the sample of stars to study is large. However if the
required memory is too large, the use of a given computer is no longer
possible or the program has to be stopped during working
hours\footnote{The computer network at Geneva Observatory links 45 machines}.

\item 
Because of the strong lines of hydrogen and calcium and the resulting 
shape of the spectrum in the range 3900 to 4100 \AA\,(see
Fig.~\ref{vega}), it is very difficult to define the continuum
there. The difficulty is enhanced by the small efficiency of ELODIE in
the blue.

\end{itemize}

Therefore the spectra were cut in the 7 parts defined in
Table~\ref{parts}.  

\begin{table}[htb]
  \caption{Definition of the 7 parts in \AA.}
  \begin{center}
    \begin{tabular}{c} \hline
      spectral range  \\
      \hline
      4125-4300  \\
      4400-4800  \\
      4950-5300  \\
      5300-5700  \\
      5700-6100  \\
      6100-6500  \\
      6580-6820  \\
      \hline\\
    \end{tabular}
    \label{parts}
  \end{center}
\end{table}

Working with 7 parts, however, implies that we have 7 different
estimates for each 
abundance. The final abundance is obtained by a weighted mean of the
7 individual estimates. Each individual value is weighted by the number of
lines having a synthetic equivalent width larger than 10 m\AA.

It is important to note that telluric lines were not corrected for before
analysis. Two of the seven parts contain a large number of these
lines. They are the two parts going from 5\,700 to 6\,500\AA. It
appears that given the width of the parts, there are enough lines for
each element so that the minimization routine is not
misled by the telluric lines. The exception may come in some slowly
rotating stars from elements with only one or two lines which are well
superposed with telluric lines.

Finally, it appears that this method is much more efficient than the
downhill simplex one when it comes to adjust abundances of 
element with only a few lines. In the first iteration, the 
  abundance of each element is adjusted in turn even if the
  $\chi^2$ value does not change a lot, whereas with the downhill
  simplex method, a step in the direction of an element with only a
  few lines
  is unlikely. Moreover, it takes less
  computer time. One reason is that a test was added to the 
  minimization routine so that an element, the abundance of which does
  not change by more than 0.1\% in two 
  successive iterations, is no longer adjusted.

\section{Validation}
To ensure that this program would work with our sample of A-F type
stars, synthetic, Vega and solar spectra were used. 
 
\subsection{Comparison with another spectrum synthesis code}\label{compsynth}

In order to check the modifications of the spectral synthesis part, a
spectrum  was produced using
\texttt{SYNSPEC} (Hubeny et al. \cite{hubeny}) with a given model
($T_\mathrm{eff} = 7\,500\mathrm{K}$, $\log{g}$ = 4.0), line list,
abundances, radial and rotational velocities (see
Table~\ref{ap}). The same input parameters were used to produce a spectrum with
our code. Both codes give almost the same spectrum as can be judged by 
eye when looking at the ratio of both spectra. That validated the
spectrum synthesis part.

In order to check the minimization routine, the spectrum
from \texttt{SYNSPEC} was used as the one to be analyzed. Since the
routine needs a starting point, solar abundances were 
used. 

\begin{table}[htb]
  \caption{Abundance determination for a synthetic spectrum computed
    with \texttt{SYNSPEC} and known abundances.}
  \begin{center}
    \begin{tabular}{c r @{.}l  r @{.} l} \hline
      Element & \multicolumn{2}{c}{Model} &
      \multicolumn{2}{c}{Adjusted}\\ 
      	& \multicolumn{2}{c}{(synspec)}& \multicolumn{2}{c}{} \\
      \hline
      $V_\mathrm{rad}~(\mathrm{km~s^{-1}})$ 	& 0	& 00	& 0&00	\\
      $V\sin{i}~(\mathrm{km~s^{-1})}$	& 15	& 00	& 15&364	\\
      $\xi_\mathrm{micro}~(\mathrm{km~s^{-1})}$	& 3	& 10  &3&00\\
      Ca		& -5	& 35	& -5&38\\
      Ti		& -7	& 00 	& -7&03\\
      V			& -7	& 30	& -7&32\\
      Cr		& -5	& 55 	& -5&58\\
      Mn		& -6	& 00 	& -6&03\\
      Fe		& -4	& 35 	& -4&37\\
      Co		& -6	& 00 	& -6&01\\
      Ni		& -5	& 89	& -5&91\\
    \hline
    \end{tabular}
    \label{ap}
  \end{center}
\end{table}

The agreement between the input and converged abundances is very good (see
Table~\ref{ap}). The difference is 
always $\leq 0.03$ dex. Moreover, all velocities
($V_\mathrm{rad}, V\sin{i}, \xi_\mathrm{micro}$) were very well
adjusted, even starting from very different values.

\subsection{Vega}\label{sec_vega}

G.M. Hill provided us with a spectrum of \object{Vega} going
from 4460 to 4530\AA, that was used to debug the
modifications of the code. 
Then a spectrum was obtained with ELODIE.
As there were a lot of changes, it is no 
longer possible to reproduce the abundances exactly as Hill's original
program,
essentially because of the change of the model atmosphere and lines
list sources. However, the abundances
estimated after the modifications are in agreement with the ones of
HL93 within 0.2 dex except for Y where only one line was used. 

For Vega, we used the model computed
especially for this star by Kurucz and available on his web page
http://cfaku5.harvard.edu/. This model is computed without convection
and with stellar parameters as follows : $T_\mathrm{eff}=9\,400$\,K,
$\log{g}=3.90$ and $\xi_{micro}=0\,\mathrm{km~s^{-1}}$

The whole procedure was run on the ELODIE spectrum and the results are 
given in Table~\ref{table_vega}.

\begin{figure*}[tbh]
\centering
 \includegraphics[width=17cm]{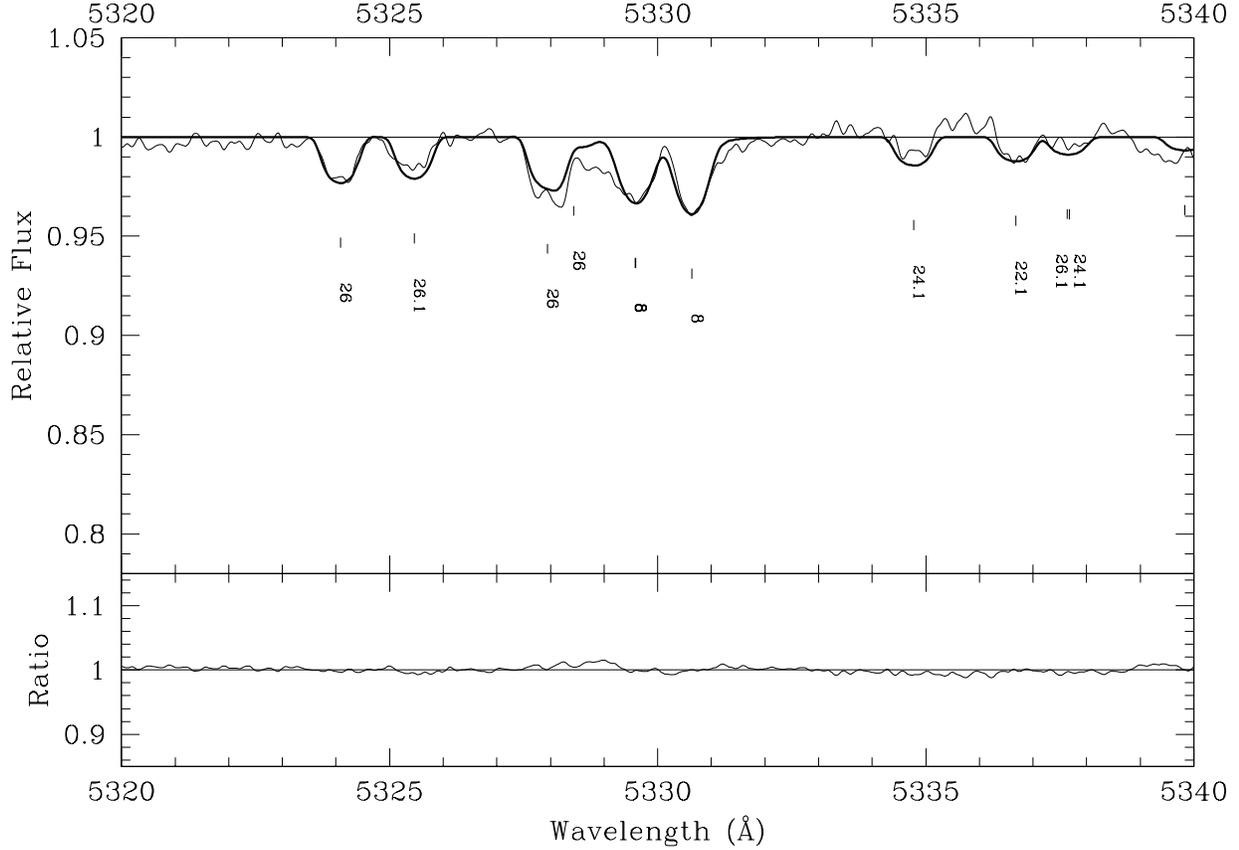}
\caption{\textbf{Top :} Superposition of a part of the observed spectrum
  (thin line)  
 and the synthetic one (thick line) for Vega. The atomic numbers and
 ionization stages (1 for singly ionized) of the species are indicated 
 under the lines. \textbf{Bottom :} Ratio
 synthetic to observed.}
\label{vega_ajust}
\end{figure*}

\begin{table}[htb]
  \caption{Derived abundances ($\log{\left[\frac{N}{N_H}\right]}$) and 
    velocities $(\mathrm{km~s^{-1}})$ for \object{Vega} compared with
      works of HL93, Adelman \& Gulliver (\cite{adelman}), Lemke
    (\cite{lemke1}, \cite{lemke2} and references therein), and Qiu et
    al. (\cite{qiu}).}
  \begin{center}
    \begin{tabular}{cccccc} \hline
      Elt & Abund & HL93 & Adelman & Lemke & Qiu \\
      \hline
      He	& -1.36 &  -1.20   & -1.52  &        &  \\
      C		& -3.51 &  -3.53   &        & -3.51  & -3.54 \\
      O 	& -3.34 &          &        &        & -2.99 \\
      Na	& -5.69 &  $<$-5.1 &        &        & -5.55 \\
      Mg	& -4.84 &  -4.69   & -5.09  &        & -5.27 \\
      Si	& -5.11 &  -5.14   &        & -5.06  & -5.04 \\
      Ca	& -6.10 &  -6.11   & -6.21  & -6.18  & -6.67 \\
      Sc	& -9.58 &          & -9.62  &        & -9.67 \\
      Ti	& -7.55 &  -7.36   & -7.47  & -7.50  & -7.42 \\
      Cr	& -6.91 &  -6.81   & -6.76  &        & -6.81 \\
      Fe	& -5.14 &  -5.03   & -5.08  & -5.03  & -5.07 \\
      Sr	& -10.03 &  $<$-7.6 &        & -9.93  & -10.72 \\
      Y 	& -9.96  &  -10.38  &        &        & -10.35 \\
      Ba	& -10.51 &  -10.51 & -10.58 & -10.57 & -11.19 \\
     \hline
      $T_\mathrm{eff}$		& 9400	 & 9560	&  9400  & 9500 & 9430\\
      $\log{g}$			& 3.95   & 4.05 &  4.03  & 3.90 & 3.95\\
      $V_\mathrm{rad}$  & -13.25  &  -13.1 & -13.26 &\\
      $V\sin{i}$        & 23.2  &  22.4  & 22.4 & \\
      $\xi_{micro}$     & 1.9   &   1.0  & 0.6    &  2.0& 1.5\\
    \hline
     \end{tabular}		
    \label{table_vega}
  \end{center}
\end{table}

\begin{figure}[tbh]
\resizebox{\hsize}{!}{\includegraphics{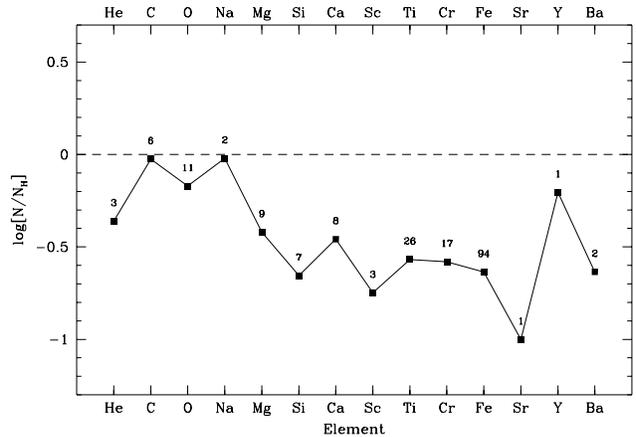}}
\caption{Logarithmic abundances of Vega with respect to the Sun. The number
indicates the number of lines with an equivalent width bigger that 10m\AA.}
\label{abund_vega}
\end{figure}

\begin{figure}[tbh]
\resizebox{\hsize}{!}{\includegraphics{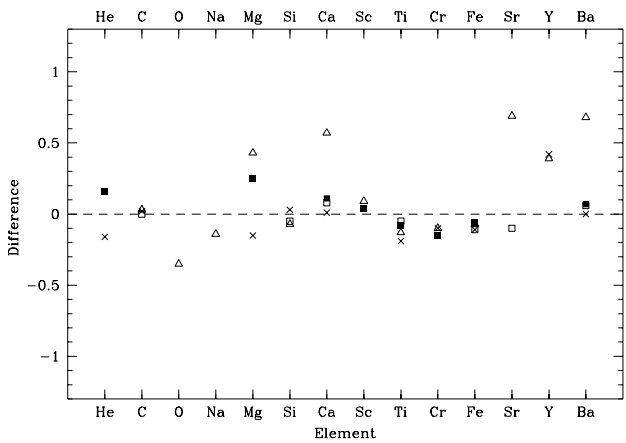}}
\caption{Difference for Vega between this paper and different
  authors ($\times$ 
HL93, $\blacksquare$ Adelman, $\square$ Lemke, and $\vartriangle$ Qiu).}
\label{diff_vega}
\end{figure}

Our estimates are in good agreement with the values available from the
literature (see Fig.~\ref{diff_vega} and
Table~\ref{table_vega}). However, it is difficult to 
compare the abundance pattern for \object{Vega} directly because of
the differences in the choice of fundamental parameters (see
Table~\ref{table_vega}). For this star, a difference of some tenths of
dex is not surprising. These differences  are the main problem when it
comes to compare results from different authors. 
Moreover, for some elements, only a few lines (sometimes only one, see 
number in Fig.~\ref{abund_vega}) are 
available and it implies that these elements are much more sensitive to 
errors on the line parameters such as $\log{gf}$. Finally, in Vega,
NLTE effects  
are not negligible for some elements. For example, a correction of 0.29 dex
for barium was calculated by Gigas (\cite{gigas}). This paper is
limited to LTE
analysis, but it will be important to check for NLTE effects when
looking for trends in element abundances.

\subsection{The Sun}\label{sect_sun}
Even if it is not easy to compare studies of various authors on
\object{Vega}, it has the advantage to be easy to
observe and it is one of the ``standard'' A0 stars even though it is
underabundant. Moreover, the original spectrum synthesis code was
intended for A-type stars so it is important to
check its validity for cooler stars since we want to use it on F-type
stars too. Therefore in order to carry out a much stronger comparison, 
a solar spectrum was used (see Sect~\ref{obs} for observation details).
 
The solar spectrum is much more crowded than the Vega one. It is not
always easy to find continuum points, and the abundance that will be
determined will show not only whether our minimization method is
efficient but also how well the continuum was
placed. In any case, a careful comparison with the Solar Atlas was
done on the whole spectral range. Fig.~\ref{compare} shows an example of the
comparison on a small spectral range.

For the Sun, we computed a model
with solar parameters ($T_\mathrm{eff}=5\,777$\,K, $\log{g}=4.44$ and
$\xi_{micro}\,=\,1\,\mathrm{km~s^{-1}}$) without overshooting.

As explained in
Sect.~\ref{list}, it was necessary to adjust some $\log{gf}$ values in order
to get ``canonical'' solar values for some elements. The biggest problem was
with Si. A lot of its lines turned out to have intensities very
different from the ones observed when computed with VALD
$\log{gf}$ (for Vega, the only useful Si lines had correct $gf$
values). Moreover, the errors were very important and could not come
from a 
wrong placement of the continuum. One can wonder why the estimated Si
abundance differs by more than 0.05 dex from the canonical one, while
$\log{gf}$ values were adjusted. In fact, we 
tried to adjust as few lines as possible. It is always possible that
small differences between observed and synthesized spectra result
from unresolved lines or weak lines that are not in the line list and
therefore not computed. A special care was brought in the computation of
lines that were not strong enough in the computed spectrum to check
how far 
a sum of weak lines might explain the gap. An interrogation of VALD around
such lines was done, showing that
the difference was never coming from forgotten lines.

\begin{figure*}[tbh]
\centering
 \includegraphics[width=17cm]{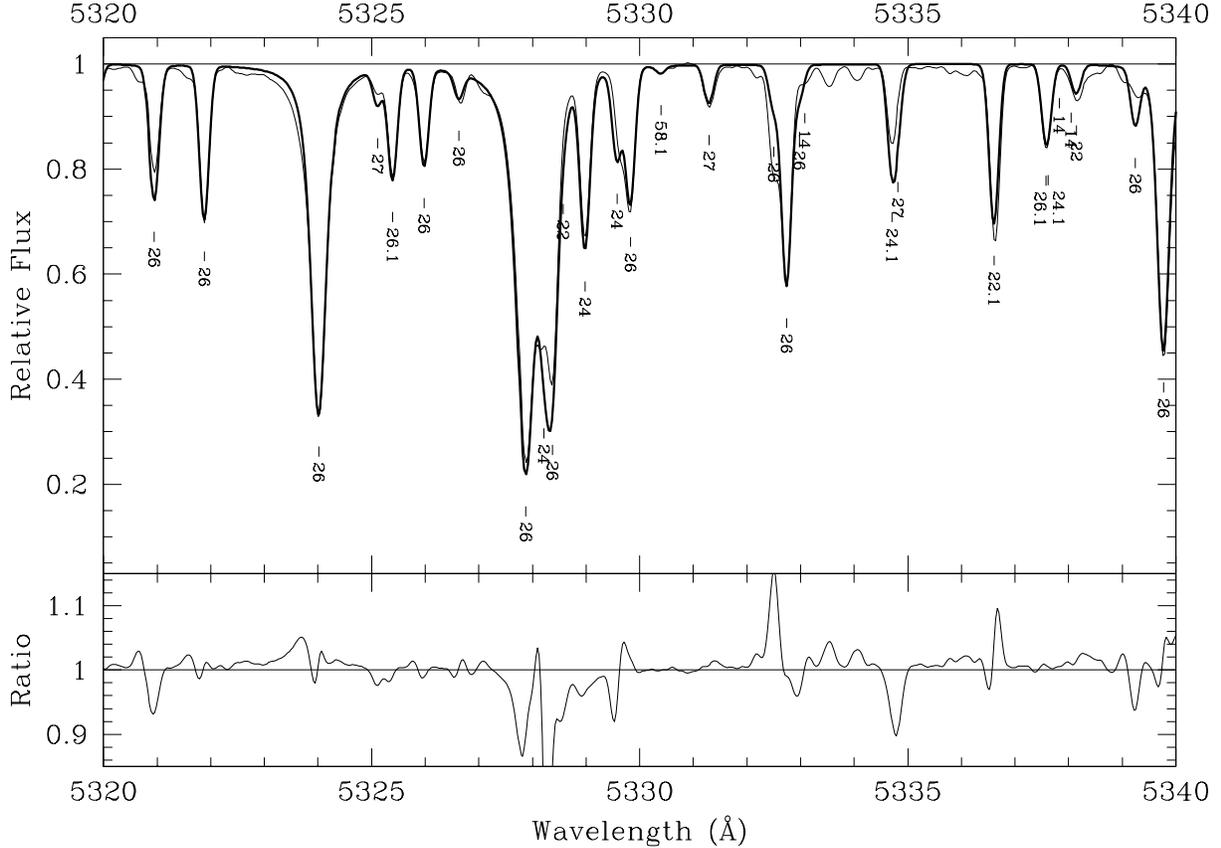}
\caption{\textbf{Top :} Superposition of a part of the observed
  spectrum (thin line)  
 and the synthetic one (thick line) for the Sun. \textbf{Bottom :} Ratio
 synthetic to observed.}
\label{sun_ajust}
\end{figure*}

\begin{table}[htb]
  \caption{Derived abundances for the \object{Sun}, difference
   with values from Grevesse and Sauval (\cite{grevesse}) and number
   of lines with an  
   equivalent width bigger than 10m\AA}
  \begin{center}
    \begin{tabular}{cccc} \hline
      Elt  &  Abundance  &  difference & \# lines\\
               &$\log{\left[\frac{N}{N_H}\right]}+12$  &  \\
    \hline
    C     &  8.56   &  0.04    & 3 \\
    Na    &  6.31   &  -0.02   & 18 \\
    Mg    &  7.52   &  -0.06   & 22 \\
    Al    &  6.42   &  -0.05   & 6 \\
    Si    &  7.48   &  -0.07   & 76 \\
    S     &  7.22   &  -0.11   & 3 \\
    Ca    &  6.34   &  -0.02   & 69 \\
    Sc    &  3.18   &  0.01    & 26 \\
    Ti    &  5.01   &  -0.01   & 361 \\
    V     &  4.04   &  0.04    &  87\\
    Cr    &  5.71   &  0.04    &  368\\
    Mn    &  5.49   &  0.10    & 81 \\
    Fe    &  7.52   &  0.02    & 1507  \\
    Co    &  4.91   &  -0.01   & 84 \\
    Ni    &  6.22   &  -0.03   & 292 \\
    Cu    &  4.23   &  0.02    & 5 \\
    Zn    &  4.69   &  0.09    & 3 \\
    Ga    &  2.84   &  -0.04   & 1 \\
    Sr    &  2.95   &  0.02    & 3 \\
    Y     &  2.20   &  -0.04   & 20 \\
    Zr    &  2.67   &  0.07    & 19\\
    Ba    &  2.15   &  0.02    & 7 \\
    La    &  1.16   &  -0.01   & 5 \\
    Ce    &  1.66   &  0.08    & 20 \\
    Nd    &  1.56   &  0.06    & 12 \\
    Sm    &  1.08   &  0.07    & 3 \\
    Eu    &  0.55   &  0.04    & 2 \\
    \hline
   \end{tabular}		
    \label{table_sun}
  \end{center}
\end{table}

\begin{figure}[tbh]
\resizebox{\hsize}{!}{\includegraphics{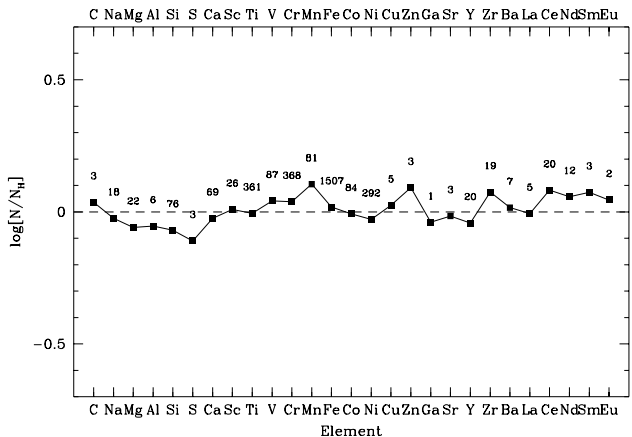}}
\caption{Same as Fig.~\ref{abund_vega}, but with the difference for the
  Sun between this paper and Grevesse \& 
  Sauval (\cite{grevesse}). }
\label{diff_sun}
\end{figure}

In the solar case, the initial abundances were chosen different from the
canonical one by some tenths of dex.
The result of our analyzis is shown in Table~\ref{table_sun} and in
Fig.~\ref{diff_sun}. A microturbulent velocity $\xi_{micro}=0.79
\,\mathrm{km~s^{-1}}$ was found, which is compatible with the value found 
by Blackwell et al. (\cite{blackwell}, $\xi_{micro}=0.775\,
\mathrm{km~s^{-1}}$) when using the model from
ATLAS9. Concerning the rotational velocities, it is important to note
that the code does not implement macroturbulence treatment. Therefore,
it is not possible to distinguish macroturbulent and rotational
velocities. A value of $3.8\,\mathrm{km~s^{-1}}$ for the ``rotational'' 
velocity was found. If we assume that the macroturbulence is
isotropic, it is possible to get a more realistic value of the
rotational velocity by doing a quadratic subtraction of the macroturbulent
velocity. Takeda (\cite{takeda1}) found that the macroturbulence change 
from 2 to 4$\,\mathrm{km~s^{-1}}$ depending of the choice of strong or
weak lines.  If we take a mean value of 3, we get $2.33
\mathrm{km~s^{-1}}$ for the rotational velocity, which is slightly
larger than the synodic value of $1.9 \mathrm{km~s^{-1}}$.

The agreement for the abundances is always better than 0.1 dex except
for S and Mn. The
difference for S results from the value of
$\log{\left[\frac{N(el)}{N_H}\right]}+12 = 7.33$ in Grevesse and Sauval
(\cite{grevesse}). However, both elements have photospheric
abundances different from the meteoritic ones by as much as 0.1 dex. The
meteoritic abundances are 7.20 and 5.53 for S and Mn
respectively. Moreover, Rentzsch-Holm (\cite{rentzsch}) found an
abundance of 7.21 for S, and in previous papers of Anders and Grevesse
(\cite{anders}), the S abundance is also 7.21, which is in perfect
agreement with our value. Finally, the line list contains only 3 weak
lines of about 15 m\AA, and therefore very sensitive to the continuum. 
Let us just stress that we do not maintain that
our value is the correct one, but that for this element, the
uncertainty is high. Concerning Mn, our value is close to the
meteoritic value too. On the other hand, hyperfine splitting can have a
significant impact and may lead to abundance overestimate
of about 0.1 dex. 

\section{Conclusion}
We have shown that our method of
determining detailed abundances for A-F type stars works and that the ELODIE
echelle spectrograph can be used to get accurate abundance
determinations for \object{Vega} 
and the \object{Sun}. We achieve a high level of automation 
to extract spectra and analyze them. One line list covering the ELODIE 
spectral range was compiled and will be used for the following study
of A-F stars. It includes new gf values for some critical lines
determined using the solar spectrum.

The problem of the choice
of stellar parameters which arose during the analysis of Vega is also a
good justification to analyze a large 
sample of stars with one given method for determining stellar
parameters ($T_\mathrm{eff}, \log{g})$. The choice between a photometric or
spectroscopic method is not so important since the uncertainty of
these methods  
is comparable. Next it is important to determine abundances 
of all stars of the sample in a homogeneous way, and
that will be possible with the method presented in this paper. This is our
final goal for which automation will be crucial. Therefore, 
even if some uncertainties remain, the resulting errors will be
systematic, and will not 
depend on the author's subjectivity.

Finally, our work is further justified by the commissioning of medium
and high resolution multi-fibers 
spectrographs, because when an observer gets hundreds of spectra each night,
he can no longer handle them by hand.

\begin{acknowledgements}
We thank Dr. G.M. Hill for providing the main program, and
for his help in 
the early stages of development
We thank Y. Chmielewski for useful and instructive discussions and for 
giving us some subroutines that proved very useful. 
We are grateful to Dr. R.O. Gray for his availability, his
kindness and the pertinence of his answers to our numerous questions.
Finally, the constructive comments of the referee,
  Dr. J. Landstreet, are gratefully acknowledged.
\end{acknowledgements}

\end{document}